\documentclass[twoside, 10pt]{article}
\usepackage[greek,american]{babel}
\usepackage{amssymb}
\usepackage{amsfonts}
\begin{document}
\newtheorem{definition}{Definition}[section]
\newtheorem{remark}{Remark}[section]
\newtheorem{defintion}{Definition}[section]
\newtheorem{theorem}{Theorem}[section]
\newtheorem{proposition}{Proposition}[section]
\newtheorem{corollary}{Corollary}[section]
\newtheorem{lemma}{Lemma}[section]
\newtheorem{help1}{Example}[section]
\title{Global existence in infinite lattices of nonlinear oscillators: The Discrete Klein-Gordon equation\footnote{\hspace*{0.5cm} AMS Subject Classification: 37L60,37L15,35Q55,34C11}}
\author{Nikos I. Karachalios\\
{\it }\\
{\it Department of Mathematics,}\\
{\it University of the Aegean,}\\
{\it Karlovassi GR 83200, Samos, GREECE}
}
\date{}
\maketitle
\pagestyle{myheadings} 
\thispagestyle{plain}
\markboth{Nikos. I. Karachalios}{Discrete Klein-Gordon equations}
\begin{abstract}
Pointing out the difference between the Discrete Nonlinear Schr\"odinger equation with the classical power law nonlinearity-for which solutions exist globally, independently of the sign and the degree of the nonlinearity, the size of the initial data and the dimension of the lattice-we prove either global existence or nonexistence in time, for the Discrete Klein-Gordon equation with the same type of nonlinearity (but of ``blow-up'' sign) , under suitable conditions on the initial data, and some times on the dimension of the lattice.  The results, consider both the conservative and the linearly damped lattice. Similarities and differences with the continuous counterparts, are remarked. We also make a short comment, on the existence of excitation thresholds, for forced solutions of damped and parametrically driven, Klein-Gordon lattices.
\end{abstract}
\section{Introduction}
Some of the most important phenomena in solid state physics (localization of electrons in disordered media, crystal dislocation), are described by inherently discrete models.  A particular example is the lattice system 
\begin{eqnarray}
\label{DW}
\ddot{\phi}_n-\epsilon(\phi_{n-1}-2\phi_n+\phi_{n+1})+m\phi_n +\nu\dot{\phi}_n+F(\phi_n)=0,
\end{eqnarray}
describing the dynamics of a one-dimensional chain of atoms, harmonically coupled to their neighbors through a parameter $\epsilon$ and subjected to a nonlinear (on-site) potential. The unknown $\phi_n(t)$ stands for the displacement of the atom $n$, while the lattice may be infinite ($n\in\mathbb{Z}$) or finite ($|n|\leq N$). The parameters $m>0$, $\nu\geq 0$ are related to the mass of the atoms and some possible linear damping effects respectively. The system (\ref{DW}) is well known as the Discrete Nonlinear Klein-Gordon equation (DKG). 

It is shown in \cite{Yuri}, that the consideration of weak linear coupling force between the particles, in the analysis of slow temporal variations of the wave envelope, retains in full the discreteness of the primary model. More precisely, in the case of a cubic and quartic potential, the above consideration associates DKG, with another famous discrete nonlinear differential equation 
\begin{eqnarray}
\label{DNLS} 
i\dot{\psi}_n+\epsilon(\psi_{n-1}-2\psi_n+\psi_{n+1})+G(\psi_n)=0,
\end{eqnarray}
$G(u_n)=\beta|\psi_n|^2\psi_n$, where $\beta$ stands for an anharmonic parameter. The lattice system (\ref{DNLS}), is known as the Discrete Nonlinear Schr\"odinger equation (DNLS). Both systems (\ref{DW}) and (\ref{DNLS}), arise in a great variety of phenomena in the context of nonlinear physics, ranging from applications in nonlinear optics to applications in biology.  The analysis of their dynamics has attracted considerable efforts, see for example \cite{Aubry,Mackayexp,Bountis,Eil2,Eil3,Kevrekidis,Yuri,Mac,Romero,Wein99} and references cited in these articles.

As a continuation of our previous works on lattice differential equations \cite{K1,AN}, here we consider the following discrete Klein-Gordon equation DKG, considered in higher dimensional lattices ($n=(n_1,n_2,\ldots,n_N)\in\mathbb{Z}^N$),
\begin{eqnarray}
\label{lat1}
\ddot{\phi}_n+\nu \dot{\phi}_n-(\Delta_d\phi)_n&+&m\phi_n +F(\phi_n)=0,\;\;n\in\mathbb{Z}^N,\;\;,t>0\\
\label{lat2}
\phi_n(0)&=&\phi_{n,0},\;\;\dot{u}_n(0)=\phi_{n,1}.
\end{eqnarray}
where 
\begin{eqnarray}
\label{DiscLap}
(\Delta_d\phi)_{n\in\mathbb{Z}^N}&=&\phi_{(n_{1}-1,n_2,\ldots ,n_N)}+\phi_{(n_1,n_{2}-1,\ldots ,n_N)}+\cdots+
\phi_{(n_1,n_{2},\cdots ,n_N-1)}\nonumber\\
&&-2N\phi_{(n_{1},n_2,\ldots ,n_N)}
+\phi_{(n_{1}+1,n_2,\ldots ,n_N)}\nonumber\\
&&+\phi_{(n_1,n_{2}+1,\ldots ,n_N)}+\cdots+
\phi_{(n_1,n_{2},\cdots ,n_N+1)}.
\end{eqnarray}
Equation (\ref{lat1}), could also be viewed as the spatial discretization of the KG partial differential equation,
\begin{eqnarray}
\label{waveeq}
\phi_{tt}+\nu {\phi}_t-\Delta\phi+m\phi +F(\phi)=0,\;x\in\mathbb{R}^N.
\end{eqnarray}
We shall consider as a model for the nonlinearity $F$, the  classical, power-law nonlinear interaction,
\vspace{.2cm}
\newline
$\mathrm{(P)}$\ \ $F(s)=\beta|s|^{2\sigma}s,\;\;s\in\mathbb{R},\;\;\beta=\pm1,\;\;\sigma>0.$
\newline

The nonlinearity $\mathrm{(P)}$ with $\beta=-1$, is usually called as a ``non-dissipative nonlinearity'' or ``blow-up'' term, since is responsible for the appearance of finite time blow-up and global nonexistence of solutions of various partial differential equations, including the KG partial differential equation. The case $\beta=1$, is sometimes called as the case of the ``dissipative'' nonlinearity. 

The motivation for the study of the lattice system (\ref{lat1})-(\ref{lat2})-$\mathrm{(P)}$  is due to the following reasons:\newline
(i) {\em Comparison with continuous counter parts.}\ As it is stated in \cite{Menzala}, ``analogy between lattices and  nonlinear partial differential equations is limited'', where the authors refer to the global existence for the DKG lattice with interaction $F(\phi_n)=\beta[(\phi_n-\phi_{n-1})^{\sigma}+ (\phi_n-\phi_{n+1})^{\sigma}]$ (FPU lattice) for $\sigma\geq 2$, and the dispersive blow-up of the long-wavelength limit, the modified KdV equation, 
$$\phi_t+\phi_{xxx}+\beta\phi^{\sigma-1}\phi_x=0,$$
which occurs for large values of $\sigma$.
In \cite{Menzala}, the authors consider the case where $\beta>0$, including nonlinearities of the form $\beta F(s)=\beta s^{p+1}$, for $p=\mathrm{odd}\geq 3$. In \cite{Menzala2}, the authors consider strongly damped lattices and moreover,  detailed decay estimates are proved, providing energy decay at exponential rate.   

Another important and characteristic example, is provided by the DNLS lattice and its continuous limit, the NLS partial differential equation. It is observed numerically \cite{bang}-we refer also to \cite{AN} for a simple proof-that solutions of conservative DNLS systems with nonlinearity $\mathrm{(P)}$ exist globally in the natural phase space, unconditionally with respect to the sign of the nonlinearity $\beta$, the degree of the nonlinearity $\sigma$,  and the size of the initial data. Sharp decay estimates of small solutions of DNLS, both for the cases $\beta\pm 1$, are rigorously proved in the recent work \cite{PK}. 

Similar behavior is shown for the the solutions of the weakly damped DNLS systems \cite{AN}: In the unforced case, it can be easily seen, that solutions of the weakly damped DNLS systems decay exponentially. For the forced system, it can be shown that a global compact attractor exists. 

This is a vast difference in comparison with the NLS-partial differential equation (\cite{cazh,cazS,Yvan} for the conservative case, \cite{Tsu1} for the weakly damped case), for which solutions may blow-up in finite time.
\newline
(ii) {\em Comparison with other lattice systems.} Our study, takes into account, the following reported observations and results: (a) the strong relation between  DNLS and DKG systems, through the analysis of slow temporal variations (b) the global existence of solutions, for the conservative or the damped DKG system for $\beta=1$ \cite{Menzala,Menzala2, PK,SZ1,SZ2} (c) the global existence results and decay estimates for small solutions of \cite{PK}, in the case $\beta=\pm1$, (d) the unconditional global existence of solutions for the conservative or damped DNLS system \cite{bang,AN, PK}.

We intend to report in this work, further possible differences and similarities between the DKG equation and the KG partial differential equation, as well as between DKG and DNLS equations, regarding the issue of nonexistence of global solutions. We also check the effect of linear dissipation, to examine if it suffices to prevent global nonexistence in time.

Close to the approach of our previous works, where we have used techniques from the theory of infinite dimensional systems, this time we are selecting some representative energy arguments \cite{Ball1b, cazh, Pohoz, Nakao93, Pucci98}, from the extensive bibliography concerning the analysis of evolution equations of second order in time. 

The preliminary section 2, is devoted to a brief description of the functional set-up, and some local existence results \cite{K1,AN}. In section 3, we consider the conservative DKG equation. We are based on the abstract framework of \cite{Pohoz}, to derive some results, on global existence and nonexistence, for negative initial energies and  initial data of definite sign. The transfer in the discrete setting of the method of \cite{Pohoz}, implies similar decay estimates and description of the interval where the solution becomes unbounded (Theorem \ref{GP}). Thus, concerning issue (i), the conservative DKG equation, exhibits similar behavior to that of the continuous counter part, since the assumptions on the initial data for either global existence and nonexistence, are similar. However, concerning the degree of the nonlinearity $\sigma$, there is not any dependence with its growth and the dimension of the lattice, as it is observed earlier, from the results on local existence (Section 2.2). This simply derived observation-as a consequence of elementary properties of the discrete phase space-seems however to be important, if viewed in comparison with the corresponding  well known restrictions for the KG partial differential equations.  The case of small data, is also treated. In this case we transfer some of the arguments of \cite{cazh}, for global existence. Let us mention that we don't claim the optimality of the derived bounds in Theorem \ref{adaptcaz} (except their interesting dependence  on the parameters $\sigma,m$), of the initial data and the global solution.  We refer for this issue, to  \cite{PK}. Especially in \cite[Theorem 8]{PK}, an  analysis using discrete analogues of Strichartz estimates is carried out, providing decay estimates, which has been tested numerically. However, taking into account their dependence on the parameters, the bounds derived by the energy argument, could be interesting in a combination with \cite[Theorem 8]{PK}, in order to check the rate of decay of the global solution obtained.

A result of global existence for the linearly damped DKG equation ($\nu>0$), is discussed in Section 4. Although the transformation $\psi=e^{-\frac{\nu}{2}t}\phi$, brings the linearly damped equation to an undamped (nonautonomous) form
making possibly applicable, the global existence results of Section 3 at least in the regime $\nu^2<<m$, we present also an alternative approach, 
based on the modified potential well method, developed in \cite{Nakao93} for the semilinear damped wave equation. 
The potential well argument is applicable through the interpolation inequality \cite[Theorem 4.1, pg.682]{Wein99},
\begin{eqnarray*}
\sum_{n\in\mathbb{Z}^N}|\phi_n|^{2\sigma+2}\leq C_*\left(\sum_{n\in\mathbb{Z}^N}|\phi_n|^2\right)^{\sigma}(-\Delta_d\phi,\phi)_{\ell^2},\;\;\sigma\geq\frac{2}{N},
\end{eqnarray*}
where $(\cdot,\cdot)_{\ell^2}$ stands for the $\ell^2$-scalar product. This inequality, which could be seen as a discrete analogue of a Sobolev-Gagliardo-Nirenberg inequality in the discrete setting, has been proved of fundamental importance for the derivation of {\em excitation thresholds} for standing wave solutions of the DNLS lattice \cite{FlachMac,Wein99}. Its application in the potential well method, reveals that the global existence and energy decay of the linearly damped DKG lattice, has a dependence under suitable assumptions on the initial data, on the nonlinearity exponent and the dimension of the lattice, through the restriction $\sigma\geq\frac{2}{N}$. Thus the result of Section 4 which is valid for multidimensional lattices, is compatible  with those of \cite{PK}, connected with the $N$-dependent excitation threshold results of \cite{Wein99}. See also Remark \ref{rem1} giving also a summary of generalizations and possible extensions.

Therefore, concerning issue (ii), the DKG equation exhibits drastically different behavior in comparison with DNLS systems, since there exist data, for which solutions cannot exist globally in time, while concerning global existence, restrictions on the size of the initial data as well as to the growth of the nonlinearity and the dimension of the lattice can appear. At this point, we mention that the fundamental difference between global nonexistence and global nonexistence which occurs by blow-up, has been analyzed in \cite{Ball1b}.

We remark that the methods are also applicable, for more general examples of nonlinearities than the model case $\mathrm{(P})$. These examples, can be generated by the assumptions of \cite{cazh, Pucci98}. Numerical testing of the results, could possibly be of some interest.  

We conclude in paragraph \ref{rem2}, with a  short comment on the existence of stationary solutions for DKG, and forced solutions for parametrically damped  and driven lattices, motivated by \cite[Section VI]{Yuri}. The comment (for the proofs we refer to \cite{K1,AN}), is related with the appearance of energy thresholds, for the existence of such type of solutions for the DKG equation, \cite{PK,Wein99}.   
\section{Preliminaries}
This introductory section is divided in two parts. In the first part, we provide some brief description of the phase space, for the infinite system of ordinary differential equations (\ref{lat1})-(\ref{lat2}). For details, we refer to \cite{AN,K1}. We also discuss some basic properties of the discrete operators defined. In the second part, we state the  result on local existence of solutions of (\ref{lat1})-(\ref{lat2}). 
\subsection{Phase space and properties of discrete operators}
We shall consider for some positive integer $N$,  the  sequence spaces denoted by
\begin{equation}
\label{ususeqs}
\ell^p=\left\{
\begin{array}{ll}
&\phi=\{\phi_n\}_{n\in\mathbb{Z}^{N}},\;n=(n_1,n_2,\ldots,n_N)\in\mathbb{Z}^N,\;\;\phi_n\in\mathbb{R},\\
&||\phi||_{\ell^p}=\left(\sum_{n\in\mathbb{Z}^N}|\phi_n|^p\right)^{\frac{1}{p}}<\infty
\end{array}
\right\}.
\end{equation}
Between $\ell^p$ spaces the following elementary embedding relation  \cite{HiLa,ree79} holds,
\begin{eqnarray}
\label{lp1}
\ell^q\subset\ell^p,\;\;\;\; ||\psi||_{\ell^p}\leq ||\psi||_{\ell^q}\,\;\; 1\leq q\leq p\leq\infty.
\end{eqnarray}
This is a contrast with the continuous analogues of $L^p(\Omega)$-spaces when $\Omega\subset\mathbb{R}^N$ has finite measure, since in this case, (\ref{lp1}) is in the opposite direction. 
The case $p=2$, stands for the usual Hilbert space of square-summable sequences, endowed with the scalar product
\begin{eqnarray}
\label{lp2}
(\phi,\psi)_{\ell^2}=\sum_{{n\in\mathbb{Z}^N}}\phi_n {\psi_n},\;\;\phi,\,\psi\in\ell^2.
\end{eqnarray}
Considering the operator $\Delta_d$ defined by (\ref{DiscLap}), we observe that for any $\phi\in\ell^2$
\begin{eqnarray}
\label{preA}
||\Delta_d\psi||_{\ell^2}^2\leq 4N||\psi||_{\ell^2}^2,
\end{eqnarray}
that is, $\Delta_d:\ell^2\rightarrow\ell^2$ is a continuous operator. Considering now the discrete operators $\nabla^+:\ell^2\rightarrow\ell^2$,  defined by
\begin{eqnarray}
\label{discder1}
(\nabla^+\psi)_{n\in\mathbb{Z}^N}&=&\left\{\psi_{(n_{1}+1,n_2,\ldots ,n_N)}-\psi_{(n_{1},n_2,\ldots ,n_N)}\right\}\nonumber\\
&+&\left\{\psi_{(n_{1},n_2+1,\ldots ,n_N)}-\psi_{(n_{1},n_2,\ldots ,n_N)}\right\}\nonumber\\
&\vdots&\nonumber\\
&+&\left\{\psi_{(n_{1},n_2,\ldots ,n_N+1)}-\psi_{(n_{1},n_2,\ldots ,n_N)}\right\},
\end{eqnarray}
and $\nabla^{-}:\ell^2\rightarrow\ell^2$ defined by
\begin{eqnarray}
\label{discder2}
(\nabla^-\psi)_{n\in\mathbb{Z}^N}&=&\left\{\psi_{(n_{1}-1,n_2,\ldots ,n_N)}-\psi_{(n_{1},n_2,\ldots ,n_N)}\right\}\nonumber\\
&+&\left\{\psi_{(n_{1},n_2-1,\ldots ,n_N)}-\psi_{(n_{1},n_2,\ldots ,n_N)}\right\}\nonumber\\
&\vdots&\nonumber\\
&+&\left\{\psi_{(n_{1},n_2,\ldots ,n_N-1)}-\psi_{(n_{1},n_2,\ldots ,n_N)}\right\},
\end{eqnarray}
and by setting
\begin{eqnarray}
\label{discder3}
(\nabla^+_{\kappa}\psi)_{n\in\mathbb{Z}^N}=\psi_{(n_{1},n_2,\ldots , n_{\kappa-1},n_{\kappa}+1,n_{\kappa+1},\ldots ,n_N)}-\psi_{(n_{1},n_2,\ldots ,n_N)},\\
\label{discder4}
(\nabla^-_{\kappa}\psi)_{n\in\mathbb{Z}^N}=\psi_{(n_{1},n_2,\ldots , n_{\kappa-1},n_{\kappa}-1,n_{\kappa+1},\ldots ,n_N)}-\psi_{(n_{1},n_2,\ldots ,n_N)},
\end{eqnarray}
we observe that the operator $\Delta_d$, satisfies the relations,
\begin{eqnarray}
\label{diffop2}
(-\Delta_d\psi_1,\psi_2)_{\ell^2}&=&\sum_{\kappa=1}^N(\nabla_\kappa^+\psi_1,\nabla_\kappa^+\psi_2)_{\ell^2},\;\;\mbox{for all}\;\;\psi_1,\psi_2\in\ell^2,\\
\label{diffop3}
(\nabla_\kappa^+\psi_1,\psi_2)_{\ell^2}&=&(\psi_1,\nabla_\kappa^-\psi_2)_{\ell^2},\;\;\mbox{for all}\;\;\psi_1,\psi_2\in\ell^2.
\end{eqnarray}
Relations (\ref{diffop2}), \ref{diffop3}), are sufficient in order to establish that  $\Delta_d:\ell^2\rightarrow\ell^2$ defines a self adjoint $\Delta_d\leq0$ operator on $\ell^2$. Examining the graph norm of the operator $\Delta_d$, we see that $D(\Delta_d)=X=\ell^2$. We shall also use the following bilinear form, and induced norm
\begin{eqnarray}
\label{lp3}
(\psi_1,\psi_2)_{\ell^2_1}&:=&\sum_{\kappa=1}^N(\nabla_\kappa^+\psi_1,\nabla_\kappa^+\psi_2)_{\ell^2}+ m(\psi_1,\psi_2)_{\ell^2},\\
\label{lp4}
||\psi||_{\ell^2_1}^2&:=&\sum_{\kappa=1}^{N}||\nabla^+_{\kappa}\psi||_{\ell_2}^2+m||\psi||_{\ell^2}^2.
\end{eqnarray}
We denote by $\ell^2_1$ the Hilbert space with  scalar product (\ref{lp3}) and norm  (\ref{lp4}). The usual norm of $\ell^2$ and (\ref{lp4}) are equivalent, since
\begin{eqnarray}
\label{lp5}
m||\psi||^2_{\ell^2}\leq ||\psi||_{\ell^2_{1}}\leq (2N+m)||\psi||^2_{\ell^2}.
\end{eqnarray}
\subsection{Local existence of solutions}
As for the DNLS lattice \cite{AN}, we shall formulate (\ref{lat1})-(\ref{lat2}), as an abstract evolution equation in $\ell^2$, \cite{Ball1b,cazh}. In the case of the DKG lattice, we may set $\mathbf{\hat{A}}:=\Delta_d-m $, and check that the operator 
\begin{equation}
\label{matrixA}
\mathbf{B}:=\left(
\begin{array}{cc}
0&1\\
\mathbf{\hat{A}}&-\nu
\end{array}
\right),\;\;D(\mathbf{B}):=\left\{\left(
\begin{array}{l}
z\\
\omega
\end{array}
\right)\,:\,z,\omega\in \ell^2_1,\,\mathbf{\hat{A}}z\in\ell^2 \right\},
\end{equation}
is a skew-adjoint operator, since 
$$
\mathbf{B}^*:=-\left(
\begin{array}{cc}
0&1\\
\mathbf{\hat{A}}&\nu
\end{array}
\right),\;\;D(\mathbf{B}^*):=\left\{\left(
\begin{array}{l}
\chi\\
\zeta
\end{array}
\right)\,:\,\chi,\zeta\in \ell^2_1,\,\mathbf{\hat{A}}\chi\in\ell^2 \right\},
$$
and $D(\mathbf{B})=D(\mathbf{B}^*)\equiv\ell_2\times\ell_2:=\ell_2^2$. The operator $\mathbf{B}$, is the generator of a an isometry group $\mathcal{T}(t):\mathbb{R}\rightarrow\mathcal{L}(\ell^2_2)$. It was shown in \cite[Lemma 2.1]{AN}, that from the nonlinear function $\mathrm{(P)}$, we may define a nonlinear map (still denoted by $F$) $F:\ell^2\rightarrow\ell^2$ for any $0\leq\sigma<\infty$. Setting
$$
\psi=\left(
\begin{array}{c}
\phi\\
\dot{\phi}
\end{array}
\right),\;\;\mathcal{F}(\psi)=\left(
\begin{array}{c}
0\\
F(\phi)
\end{array}
\right),
$$
equation (\ref{lat1}) can be rewritten as
\begin{eqnarray}
\label{system}
\dot{\psi}=\mathbf{B}\psi+\mathcal{F}(\psi).
\end{eqnarray}
For fixed $T>0$ and $(\{\phi_{n,0}\}_{n\in\mathbb{Z}},\{\phi_{n,1}\}_{n\in\mathbb{Z}})^{T}=(\phi_0, \phi_1)^T:=\psi_0\in\ell^2_2$, a function $\psi:=(\phi,\dot{\phi})\in\mathrm{C}([0,T],\ell^2_2)$ is a solution of (\ref{system}), if and only if
\begin{eqnarray}
\label{milds}
\psi(t)=\mathcal{T}(t)\psi_0+\int_{0}^{t}\mathcal{T}(t-s)\mathcal{F}(\psi(s))ds:=S(t)\psi_0,\;\;
\end{eqnarray}
Our local existence result can be stated as follows:
\begin{proposition}
\label{locex}
We consider the DKG (\ref{lat1})-(\ref{lat2})-$\mathrm{(P)}$, with $\nu\geq 0$. We assume that $\psi_0\in\ell^2_2$. Then, there exists a function $T^*:\ell^2_2\rightarrow (0,\infty]$ with the following properties:\vspace{.2cm}\\
(a)\  For all $\psi_0\in\ell^2_2$, there exists  $\psi\in\mathrm{C}([0,T^*(\psi_0)),\ell^2_2)$ such that
for all $0<T<T^*(\psi_0)$, $\psi$ is the unique solution of (\ref{system}) in $\mathrm{C}([0,T],\ell^2_2)$ (well posedness).\vspace{.2cm}\\
(b)\  For all $t\in [0,T^*(\psi_0))$,
\begin{eqnarray}
\label{maxT}
T^*(\psi_0)-t\geq \frac{1}{2(L(R)+1)}:=T_R,\;\;R=2||\psi(t)||_{\ell^2_2},
\end{eqnarray}
where $L$ denotes the Lipschitz constant associated with the nonlinear operator $\mathcal{F}:\ell^2_2\rightarrow\ell^2_2$. Moreover the following alternative holds: (i) $T^{*}(\psi_0)=\infty$, or (ii) $T^*(\psi_0)<\infty$ and $$\lim_{T\uparrow T^*(\psi_0)}||\psi(t)||_{\ell^2_2}=\infty,\;\;\mbox{(maximality)}.$$
(c)\ $T^*:\ell^2_2\rightarrow (0,\infty]$ is lower semicontinuous.
In addition, if $\{\psi_{n0}\}_{n\in\mathbb{N}}$ is a sequence in
$\ell^2_2$ such that $\psi_{n0}\rightarrow \psi_0$ and if $T<T^*(\psi_0)$,
then $S(t)\psi_{0n}\rightarrow S(t)\psi_0$ in $\mathrm{C}([0,T],\ell^2)$
(continuous dependence on initial data).
\end{proposition}

It is important to remark further continuity properties of the solution, which will be essential for the application of the energy arguments, in the next sections. We have the following 
\begin{corollary}
\label{contnorms}
The solution of (\ref{lat1})-(\ref{lat2})-$\mathrm{(P)}$, is such that $\phi\in \mathrm{C}^2([0,T];\ell^2)\cap \mathrm{C}([0,T];\ell^{2\sigma +2})$ for all $0<\sigma\leq\infty$.
\end{corollary}
{\bf Proof:}\ 
It follows from Proposition \ref{locex}, equation (\ref{lat1}), the continuity of the operator $\Delta_d:\ell^2\rightarrow\ell^2$ and the Lipschitz continuity of the nonlinear map $F:\ell^2\rightarrow\ell^2$, that solutions $\phi$ of (\ref{lat1})-(\ref{lat2})-$\mathrm{(P)}$, are such that $\phi\in \mathrm{C}^2([0,T];\ell^2)$, since
$$\ddot{\phi}_n+\nu \dot{\phi}_n=(\Delta_d\phi)_n-m\phi_n -F(\phi_n)=0,\;\;n\in\mathbb{Z}^N,\;\;,t\in [0,T].$$
From the embedding (\ref{lp1}), we have that $\ell^2\subset\ell^{2\sigma+2}$ for any $0<\sigma\leq\infty$ and \cite[Vol. II/A, Proposition 23.2 (c)\& (h), pg. 407]{zei85}, implies that $L^{r_1}([0,T];\ell^2)\subseteq L^{r_2}([0,T];\ell^{2\sigma+2})$ for all
$1\leq r_2\leq r_1<\infty$.  Therefore
$$\phi\in\mathcal{W}:=\left\{\phi\,:\,\phi\in L^{p}([0,T];\ell^2),\,\dot{\phi}\in L^{q}([0,T];\ell^{2\sigma+2})\right\}\;\;1\leq p,q<\infty.$$ 
Then from \cite[Vol. II/A, 23.13a, pg. 450]{zei85}, we conclude that $\phi\in \mathrm{C}([0,T];\ell^{2\sigma +2}).$\ \ $\diamond$ 
\section{The conservative DKG equation} 
In this section, we discuss the global in time solvability of the conservative DKG equation (\ref{lat1})-(\ref{lat2}) ($\nu=0$). Our analysis is based on the study of the total energy functional
\begin{eqnarray*} 
\label{Gal10}
\mathcal{H}(t)=\frac{1}{2}||\dot{\phi}(t)||_{\ell^2}^2+\frac{1}{2}\left\{\sum_{\kappa=1}^{N}||\nabla^+_{\kappa}\phi(t)||_{\ell_2}^2+m||\phi||_{\ell^2}^2\right\}
-\frac{1}{2\sigma +2}\sum_{n\in\mathbb{Z}^N}|\phi_n(t)|^{2\sigma+2}.
\end{eqnarray*}
The functional $\mathcal{H}$, is the Hamiltonian of the infinite lattice of nonlinear oscillators: it is conserved as long as the unique solution of (\ref{lat1})-(\ref{lat2}) exists, that is
\begin{eqnarray}
\label{Gal1}
\mathcal{H}(\phi_0,\phi_1):=\mathcal{H}(0)=\mathcal{H}(t),\;\;t\in[0,T],\;\;\mbox{for all}\;\;0<T<T^*.
\end{eqnarray}
\paragraph{The case of non-positive initial energy.}
We discuss first, the issue of global nonexistence of solutions.  We will show that the abstract framework developed  in \cite{Pohoz}, can be adapted for the study of the DKG (\ref{lat1})-(\ref{lat2})-$\mathrm{(P)}$.  
The differentiability of the functional
\begin{eqnarray}
\label{Gal2}
\mathcal{V}(\phi)=-\frac{1}{2}\left\{\sum_{\kappa=1}^{N}||\nabla^+_{\kappa}\phi(t)||_{\ell_2}^2+m||\phi||_{\ell^2}^2\right\}+\frac{1}{2\sigma+2}\sum_{n\in\mathbb{Z}^N}|\phi_n(t)|^{2\sigma+2}.
\end{eqnarray}
is needed. This can be done by using (\ref{lp1}), and working exactly as in \cite[Lemma 2.2]{K1}. The proof makes use of a discrete analogue of dominated convergence theorem \cite{Bates}.
\begin{lemma}
\label{aux1}
Let $\phi\in\ell^{2}$ and $0<\sigma <\infty$. Then the functional (\ref{Gal2})
is a $\mathrm{C}^{1}(\ell^{2},\mathbb{R})$ functional and
\begin{eqnarray}
\label{gatdev}
<\mathcal{V}'(\phi),\psi>=&-&\left\{\sum_{\kappa=1}^N\sum_{n\in\mathbb{Z}^N}(\nabla^+_{\kappa}\phi)_{n}(\nabla^+_{\kappa}\psi)_{n}+m\sum_{n\in\mathbb{Z}^N}\phi_n\psi_n\right\}\nonumber\\
&+&\sum_{n\in\mathbb{Z}^N}|\phi_n|^{2\sigma}\phi_n{\psi_n},\;\;\psi\in\ell^{2}.
\end{eqnarray}
\end{lemma}

We seek for some $\lambda >2$, satisfying the key condition of  \cite[Inequality (2.6), pg. 455]{Pohoz}. Using (\ref{Gal2}) and (\ref{gatdev}), the definition of the equivalent norm $\ell^2_1$ (\ref{lp4}), (\ref{lp5}) and (\ref{lp1}), we get that 
\begin{eqnarray}
\label{Gal4}
 \left<\mathcal{V}'(\phi),\phi\right>-\lambda \mathcal{V}(\phi)=\left(\frac{\lambda}{2}-1\right)||\phi||^2_{\ell^2_1}+\left(1-\frac{\lambda}{2\sigma +2}\right)||\phi||_{\ell^{2\sigma+2}}^{2\sigma+2},
\end{eqnarray}
Setting $\lambda=2\sigma+2$, we observe that 
\begin{eqnarray}
\label{Gal5}
\left<\mathcal{V}'(\phi),\phi\right>-\lambda \mathcal{V}(\phi)\geq \sigma ||\phi||^2_{\ell^2_1}\geq 0.
\end{eqnarray}
Using Corollary \ref{contnorms}, for the time derivatives of $\mu(t)=||\phi(t)||^2_{\ell^2}$, we have the relations
\begin{eqnarray}
\label{Gal6}
\mu '(t)&=&2(\phi,\dot{\phi})_{\ell^2},\nonumber\\
\mu ''(t)&=&2||\dot{\phi}||^2_{\ell^2}+2\left<\mathcal{V}'(\phi),\phi\right>
\geq 2||\dot{\phi}||^2_{\ell^2}+4(\sigma +1)\mathcal{V}(\phi).
\end{eqnarray}
which can be derived by using (\ref{lat1}), the  Hamiltonian $\mathcal{H}$ and (\ref{Gal2}). We have 
$$\mu'(0)=2(\phi_0,\phi_1)_{\ell^2},\;\;\mu(0)=||\phi_0||^2_{\ell^2}.$$
We assume nonpositive initial energy
\begin{eqnarray}
\label{Gal7}
\mathcal{H}(0)\leq 0.
\end{eqnarray}
Now, from the conservation of energy (\ref{Gal1}) and (\ref{Gal6}), (\ref{Gal7}), we get
\begin{eqnarray}
\label{Gal8}
\mu ''(t)\geq 2||\dot{\phi}||_{\ell^2}^2+4(\sigma +1)\left(\frac{1}{2}||\dot{\phi}||_{\ell^2}^2-\mathcal{H}(0)\right)\geq (2\sigma +4)||\dot{\phi}||^2_{\ell^2}.
\end{eqnarray}
On the other hand we have that
\begin{eqnarray}
\label{Gal8a} 
(\mu'(t))^2\leq 4\mu(t)^2||\dot{\phi}||^2_{\ell^2}. 
\end{eqnarray}
The differential inequality  
\begin{eqnarray}
\label{Gal9}
\mu''(t)\geq \frac{\sigma+2}{2}\frac{[\mu'(t)]^2}{\mu(t)},\;\;t>0,
\end{eqnarray}
which can be written also in the form
\begin{eqnarray}
\label{Gal9b}
\frac{\mu''(t)}{[\mu'(t)]^2}\geq \frac{\sigma+2}{2}\frac{1}{\mu(t)},
\end{eqnarray}
is derived by combination of (\ref{Gal8a}) and(\ref{Gal8}). As in \cite[pg. 455]{Pohoz}, we may conclude with the following observations:\newline
Assuming that $\mu'(0)>0$, by integrating (\ref{Gal9}), we obtain the Ricatti inequality
\begin{eqnarray}
\label{Ric}
\frac{\mu'(t)}{[\mu(t)]^{\frac{\sigma +2}{2}}}\geq\frac{\mu'(0)}{[\mu(0)]^{\frac{\sigma +2}{2}}}>0,
\end{eqnarray}
with blow-up time estimated by 
$$T^*<\frac{2}{\sigma}\frac{\mu(0)}{\mu'(0)}.$$
On the other hand, assuming that $\mu'(0)<0$,  by integration of (\ref{Gal9b}), we get that
\begin{eqnarray*}
[\mu(t)]^{-1}\leq[\mu'(0)]^{-1}- \frac{\sigma+2}{2}\int_0^t[\mu(s)]^{-1}ds,
\end{eqnarray*}
implying that $\mu'(t)<0$, for all $t>0$. We summarize in the following theorem, whose proof is identical to those of
\cite[Lemma 2.1 \& Lemma 2.2]{Pohoz}. 
\begin{theorem}
\label{GP}
We consider the DKG equation (\ref{lat1})-(\ref{lat2})-$\mathrm{(P)}$, with $\beta=-1$ and $0<\sigma<+\infty$. 
(i)\ We assume that the initial data are such that $\mathcal{H}(0)\leq 0$ and $(\phi_0,\phi_1)_{\ell^2}<0$. Then the unique solution $\phi\in C^2([0,+\infty),\ell^2)$, and satisfies the decay estimate
\begin{eqnarray*}
\label{rate1}
||\phi(t)||_{\ell^2}\leq ||\phi_0||_{\ell^2}(1+\delta t)^{-\frac{1}{\sigma}},\;\;\delta=\frac{\sigma|(\phi_0,\phi_1)_{\ell^2}|}{||\phi_0||^2}.
\end{eqnarray*}
(ii)\ We assume that the initial data are such that $\mathcal{H}(0)\leq 0$ and $(\phi_0,\phi_1)_{\ell^2}>0$. Then the solution blows up (in the sense that $\mu(t)=||u(t)||^2_{\ell^2}$ becomes unbounded) on the finite interval $(0,T^*)$, with 
\begin{eqnarray}
\label{estbl}
T^*=\frac{||\phi_0||^2}{\sigma(\phi_0,\phi_1)_{\ell^2}}.
\end{eqnarray}
\end{theorem}
\paragraph{Global existence results  for sufficiently small initial data.}
In the present paragraph, we show that solutions still exist globally in time, if we assume conditions only on the size of the norm of the initial data. This time, we choose to provide a ``discrete version'' of \cite[Proposition 6.3.3]{cazh}. Inspection of the proof, will reveal that discreteness may have some additional interesting effects, in comparison with the continuous case (see Remark \ref{rem1}). The proof could be viewed as a simple alternative to \cite[Theorem 8]{PK}. The strength of \cite[Theorem 8]{PK}, is also on the detailed decay estimates of the solution and its connection with excitation thresholds, for the existence of standing waves (see ).  However, the conditions needed for the result presented here, seem to be consistent with the excitation thresholds of paragraph \ref{rem2}, since there are also independent of the dimension of the lattice and the growth of the nonlinearity exponent.
\begin{theorem}
\label{adaptcaz}
We consider the DKG equation (\ref{lat1})-(\ref{lat2})-$\mathrm{(P)}$, with $\beta=-1$ and $0<\sigma<+\infty$. We consider also the function
\begin{eqnarray}
\label{gro9}
\theta(x)=M_0x^{1+\sigma}-x,\;\;x\in [0,+\infty),\;\;M_0=\frac{1}{m^{\sigma+1}(\sigma +1)},
\end{eqnarray}
and we set $-\rho:=\min\theta(x)<0$.
Assume that the initial data $\psi_0=(\phi_0, \phi_1)^T\in\ell^2_2$ are sufficiently small in the sense 
\begin{eqnarray}
\label{eil1}
||\psi_0||_1^2:=||{\phi}_1||_{\ell^2}^2+||\phi_0||_{\ell^2_1}^2\leq\min\left\{1,\frac{\rho m^{\sigma +1}(\sigma+1)}{1+m^{\sigma +1}(\sigma+1)}\right\},
\end{eqnarray}
Then $\phi\in C^2([0,+\infty))$, and satisfies the estimate
\begin{eqnarray}
\label{eil2}
\sup_{t\geq 0}||\dot{\phi}(t)||_{\ell^2}^2+||\phi(t)||_{\ell^2_1}^2\leq 
\frac{m^{\sigma +1}(\sigma+1)^2}{(1+m^{\sigma +1})(\sigma+1)\sigma}||\psi_0||_1^2.
\end{eqnarray}
\end{theorem}
{\bf Proof:} We shall consider the function
\begin{eqnarray}
\label{gro4}
\mu(t)=||\dot{\phi}(t)||_{\ell^2}^2+||\phi(t)||_{\ell^2_1}^2.
\end{eqnarray}
It is convenient to rewrite the conservation of energy (\ref{Gal1}), as 
\begin{eqnarray}
\label{gro5}
\mu(t)=\mu(0)-2\mathbf{J}(\phi_0)+2\mathbf{J}(\phi(t)),\;\;\mathbf{J}(\phi(t))=\frac{1}{2\sigma +2}\sum_{n\in\mathbb{Z}^N}|\phi_n(t)|^{2\sigma+2}.
\end{eqnarray}
By using the embedding relation (\ref{lp1}), and the equivalence of norms (\ref{lp5}), we easily derive the inequality
\begin{eqnarray}
\label{gro6}
2\mathbf{J}(\phi(t))=\frac{1}{\sigma +1}||\phi(t)||_{\ell_{2\sigma +2}}^{2\sigma +2}&\leq& \frac{1}{\sigma +1}||\phi(t)||_{\ell^2}^{2\sigma +2}\nonumber\\ 
&\leq&
\frac{1}{(\sigma +1)m^{\sigma +1}}||\phi||_{\ell^2_1}^{2\sigma +2}\nonumber\\
&\leq& \frac{1}{(\sigma +1)m^{\sigma +1}}\mu (t)^{\sigma +1}.
\end{eqnarray}
Thus, from (\ref{gro5}) and (\ref{gro6}) we get that
\begin{eqnarray}
\label{gro6a}
\mu(t)\leq \mu(0)-2\mathbf{J}(\phi_0)+\frac{1}{(\sigma +1)m^{\sigma +1}}\mu (t)^{\sigma +1}.
\end{eqnarray}
The same arguments ((\ref{gro6}) for $\phi_0$) can be applied, in order to show that inequality (\ref{gro6a}), holds for the initial value $\mu(0)$. We have that
\begin{eqnarray}
\label{gro7}
\mu(0)-2\mathbf{J}(\phi_0)\leq \mu(0)+\frac{1}{(\sigma+1)m^{\sigma +1}}\mu (0)^{\sigma +1}.
\end{eqnarray} 
A first assumption on the size of the initial data is 
\begin{eqnarray}
\label{small1}
\mu(0)<1. 
\end{eqnarray}
Then, from (\ref{gro7}) it follows that
\begin{eqnarray}
\label{gro8}
\mu(0)-2\mathbf{J}(\phi_0)\leq M_1\mu(0),\;\;M_1=1+M_0.
\end{eqnarray}
The rest, will follow by repeating the arguments of \cite[Proposition 6.3.3, pg. 85-86]{cazh}: The consideration of the function $\theta(x)$, is motivated by inequality (\ref{gro6a}). It appears that for all $\alpha\in (0,\rho)$, there exist $x_{\alpha},y_{\alpha}$, such that $0<x_{\alpha}<y_{\alpha}$, satisfying $\theta(x_{\alpha})+\alpha=\theta(y_{\alpha})+\alpha=0$. Moreover it holds that $\alpha<x_{\alpha}<\alpha (1+\sigma)/\sigma$. Combining (\ref{gro6a}) and (\ref{gro8}), we observe that
\begin{eqnarray*}
\theta(\mu(t))+M_1\mu (0)\geq 0,
\end{eqnarray*}
for all $t\in [0,T]$. Thus, assuming that 
\begin{eqnarray}
\label{caz1}
M_1\mu(0)<\rho,
\end{eqnarray} 
we may set $\alpha=M_1\mu(0)$, in order to apply the aforementioned argument, for the function $\theta(x)$ and the point $x_{\alpha}$. We obtain that
\begin{eqnarray}
\label{gro10}
\mu(t)\in [0,x_{M_1\mu(0)})\cup (y_{M_1\mu(0)},+\infty).
\end{eqnarray}
Now the continuity of the function $\mu(t)$, implies that
\begin{eqnarray}
\label{caz2}
\mu(t)\leq x_{M_1\mu(0)}\leq\frac{1+\sigma}{\sigma}M_1\mu (0),
\end{eqnarray}
for all $t\in [0,T)$. Inequalities (\ref{small1}) and (\ref{caz1}), enforce us to choose $\mu(0)<\min (1,\rho M_1^{-1})$, hence (\ref{eil1}) on the size of the norm of the initial data.  The last part of the rhs of inequality (\ref{caz2}),  is the estimate (\ref{eil2}) on the corresponding global solution.\ \ $\diamond$ 

\section{The linearly damped DKG} 
In this section, we consider the linearly damped DKG equation (\ref{lat1})-(\ref{lat2})-$\mathrm{(P)}$.  Let us note that the transformation $\psi\rightarrow e^{-\frac{\nu}{2}t}\phi$, brings the equation (\ref{lat1}), into the nonautonomous undamped DKG equation
\begin{eqnarray*}
\ddot{\phi}_n&-&(\Delta_d\phi)_n+\tilde{m}_{\nu}\phi_n+\beta F(\phi_n,t)=0,\;\;n\in\mathbb{Z}^N,
\end{eqnarray*}
where $\tilde{m}_{\nu}=m-\frac{\nu^2}{2}$ and $F(t,\phi_n)=e^{-\nu\sigma t}|\phi_n|^{2\sigma}\phi_n$. Theorem \ref{adaptcaz} seems to be applicable (after some modifications) at least in the regime $m>>\nu^2$, which is a physically interesting case. 
However, we decide to  present a global existence and energy decay result, for the equation 
\begin{eqnarray}
\label{lat1m}
\ddot{\phi}_n+\nu \dot{\phi}_n&-&(\Delta_d\phi)_n+\beta|\phi_n|^{2\sigma}\phi_n=0,\;\;n\in\mathbb{Z}^N,\;\;t>0,\\
\label{lat2m}
\phi_n(0)&=&\phi_{n,0},\;\;\dot{\phi}_n(0)=\phi_{n,1}.
\end{eqnarray} 
based on the the ``modified potential well'', introduced in \cite{Nakao93} for the study of the linearly damped semilinear wave equation. The result makes use of a discrete analogue of the Sobolev-Gagliardo-Nirenberg inequality, which was proved in \cite{Wein99}, in the context of existence of excitation thresholds for the DNLS equation with power nonlinearity. The potential well arguments have their origin in \cite{Payne75}. The result will demonstrate that under certain assumptions on the size of the initial data, the system features global existence and energy decay, assuming sufficient growth of the nonlinearity, depending on the dimension. A detailed energy decay estimate, is also derived, see (\ref{energydecay})-(\ref{1stcon})-(\ref{2ndcon}). 

We recall first, the discrete interpolation inequality, which was proved in \cite{Wein99}.
\begin{theorem}
\label{WGN}
Assume that $\sigma\geq\frac{2}{N}$. Then there exists $C^*>0$, such that for all $\phi\in\ell^2$, holds the interpolation inequality
\begin{eqnarray}
\label{WGN1}
\sum_{n\in\mathbb{Z}^N}|\phi_n|^{2\sigma+2}\leq C_*\left(\sum_{n\in\mathbb{Z}^N}|\phi_n|^2\right)^{\sigma}\sum_{\kappa=1}^{N}||\nabla_k^+\phi||^2_{\ell^2}.
\end{eqnarray}
\end{theorem}

We rewrite for convenience in the case $m=0$, the energy quantities
\begin{eqnarray}
\label{mf1}
\mathcal{H}(t)&=&\frac{1}{2}||\dot{\phi}(t)||_{\ell^2}^2+\frac{1}{2}\sum_{\kappa=1}^{N}||\nabla_k^+\phi(t)||^2_{\ell^2}
-\frac{1}{2\sigma +2}\sum_{n\in\mathbb{Z}^N}|\phi_n(t)|^{2\sigma+2},\\
\label{mf2}
\mathcal{V}(\phi(t))&=&-\frac{1}{2}\sum_{\kappa=1}^{N}||\nabla_k^+\phi(t)||^2_{\ell^2}+\frac{1}{2\sigma+2}\sum_{n\in\mathbb{Z}^N}|\phi_n(t)|^{2\sigma+2}.
\end{eqnarray}
Multiplying equation (\ref{lat1m}), by $\dot{\phi}$ in the $\ell^2$- scalar product, and by using (\ref{diffop2}), we derive the relation of dissipation of energy
\begin{eqnarray}
\label{dissE}
\mathcal{H}(t)=\mathcal{H}(0)-\nu\int_{0}^{t}||\dot{\phi}(s)||^2_{\ell^2}ds,
\end{eqnarray}
which hold as long as the local solution exists. Moreover, we observe that
\begin{eqnarray}
\label{TvsP}
-\mathcal{V}(t)\leq \mathcal{H}(t).
\end{eqnarray}

With the help of the inequality (\ref{WGN1}), we shall proceed to the proof of  

\begin{theorem}
\label{Nakono}
Let  $\beta=-1$, in in (\ref{lat1m}), and we assume that the nonlinearity exponent satisfies
$\sigma\geq\frac{2}{N}$. Setting
$$\rho_0:=\frac{2}{\nu}\left\{\nu||\phi_0||^2_{\ell^2}+\frac{6}{\nu}\mathcal{H}(0)+2(\phi_0,\phi_1)_{\ell^2}\right\},$$
we assume that the initial data (\ref{lat2m}), satisfy the conditions
\begin{eqnarray}
\label{glona}
\rho_0\leq\frac{1}{C_*^{\sigma}}\;\;\mbox{and}\;\;
\sum_{\kappa=1}^{N}||\nabla_k^+\phi_0||^2_{\ell^2}>||\phi_0||_{\ell^{2\sigma +2}}^{2\sigma +2}.
\end{eqnarray}
Then the solution of (\ref{lat1m})-(\ref{lat2m}) $\phi\in C^2([0,+\infty),\ell^2)$, and $\lim_{t\rightarrow +\infty}||\phi(t)||_{\ell^2}=0$.
\end{theorem}
{\bf Proof:} 
The modified potential well in our case, will be defined as
\begin{eqnarray}
\label{discmod}
\mathrm{W}_{\ell^2}:=\mbox{the interior of the set}\;\left\{\phi\in\ell^2|K(\phi)=\sum_{\kappa=1}^{N}||\nabla_k^+\phi||^2_{\ell^2}-||\phi||_{\ell^{2\sigma+2}}^{2\sigma+2}\geq 0\right\}.
\end{eqnarray}
To the fact that $\mathrm{W}_{\ell^2}$ is well defined for $\sigma\geq\frac{2}{N}$, is a consequence of the discrete interpolation inequality (\ref{WGN1}) (see also \cite[pg. 329, (1.13)]{Nakao93}, for the continuum case).

We argue  by contradiction. Let us assume that there exists $T^*>0$, such that 
\begin{eqnarray*}
\phi(t)\in \mathrm{W}_{\ell^2}\;\;\mbox{for all}\;\;t\in[0,T^*),\;\;\mbox{and}\;\;\phi(T^*)\in\partial\mathrm{W}_{\ell^2}.
\end{eqnarray*}
Then it follows that $K(\phi(T^*))=0$, and $\phi(T^*)\neq 0$. For every $t\in [0,T^*]$, it follows from (\ref{discmod}), that
\begin{eqnarray}
\label{es1}
-\mathcal{V}(t)=\frac{1}{2}\sum_{\kappa=1}^{N}||\nabla_k^+\phi(t)||^2_{\ell^2}-\frac{1}{2\sigma+2}||\phi(t)||_{\ell^{2\sigma+2}}^{2\sigma+2}
\geq\frac{\sigma}{2\sigma+2}\sum_{\kappa=1}^{N}||\nabla_k^+\phi(t)||^2_{\ell^2}.
\end{eqnarray}
Multiplication of (\ref{lat1m}), by $\phi$ in the $\ell^2$ scalar product, implies the identity 
\begin{eqnarray}
\label{blo9}
2\frac{d}{dt} (\phi(t),\dot{\phi}(t))_{\ell^2}-2||\dot{\phi}(t)||_{\ell^2}^2+\nu\frac{d}{dt}||\phi(t)||_{\ell^2}^2
+2\sum_{\kappa=1}^{N}||\nabla^+_{\kappa}\phi(t)||_{\ell_2}=2||\phi(t)||_{\ell^{2\sigma+2}}^{2\sigma+2}.
\end{eqnarray}
Since $K(\phi(t)\geq 0$, for all $0\leq t\leq T^*$, we obtain from (\ref{blo9}), that
\begin{eqnarray}
\label{blo91}
\nu||\phi(t)||^2_{\ell^2}&\leq&\nu||\phi_0||_{\ell^2}^2+2|(\phi(t),\dot{\phi}(t))_{\ell^2}|+2(\phi_0,\phi_1)_{\ell^2}
+2\int_{0}^t||\dot{\phi}(s)||_{\ell^2}^2ds\nonumber\\
&\leq&
\nu||\phi_0||_{\ell^2}^2+2\left\{\frac{\nu}{4}||\phi(t)||_{\ell^2}^2+\frac{1}{\nu}||\dot{\phi}(t)||_{\ell^2}^2\right\}
+2(\phi_0,\phi_1)_{\ell^2}+2\int_{0}^t||\dot{\phi}(s)||_{\ell^2}^2ds,
\end{eqnarray}
Moreover, from (\ref{es1}), we have that
\begin{eqnarray}
\label{blo92}
\mathcal{H}(t)=\frac{1}{2}||\dot{\phi}(t)||_{\ell^2}^2-\mathcal{V}(t)\geq \frac{1}{2}||\dot{\phi}(t)||_{\ell^2}^2+
\frac{\sigma}{2\sigma+2}\sum_{\kappa=1}^{N}||\nabla_k^+\phi(t)||^2_{\ell^2}\geq \frac{1}{2}||\dot{\phi}(t)||_{\ell^2}^2,
\end{eqnarray}
for all $0\leq t\leq T^*$. 
Now, from the inequality (\ref{blo92}) and the identity (\ref{dissE}), 
we get the estimate
\begin{eqnarray}
\label{blo93}
\frac{1}{2}||\dot{\phi}(t)||_{\ell^2}^2+\nu\int_{0}^t||\dot{\phi}(s)||_{\ell^2}^2ds\leq\mathcal{H}(0).
\end{eqnarray}
Then, inserting (\ref{blo93}) into (\ref{blo91}), we get that
\begin{eqnarray}
\label{blo94}
||\phi(t)||^2_{\ell^2}\leq \frac{2}{\nu}\left\{\nu||\phi_0||^2_{\ell^2}+\frac{6}{\nu}\mathcal{H}(0)+2(\phi_0,\phi_1)_{\ell^2}\right\}:=\rho_0^2
\end{eqnarray}
for all $0\leq t\leq T^*$. With the bound (\ref{blo94}) at hand, we use the inequality (\ref{WGN1}), to obtain
\begin{eqnarray}
\label{blo95}
\sum_{n\in\mathbb{Z}^N}|\phi_n(t)|^{2\sigma+2}\leq C_*\rho_0^{\sigma}\sum_{\kappa=1}^{N}||\nabla_k^+\phi(t)||^2_{\ell^2},\;\;0\leq t\leq T^*.
\end{eqnarray}
We set
$$\lambda_*:=C_*\rho_0^\sigma.$$
Hence, by using (\ref{blo95}), we derive that for $t=T^*$,
\begin{eqnarray*}
K(\phi(T^*))&=&\sum_{\kappa=1}^{N}||\nabla_k^+\phi(T^*)||^2_{\ell^2}-||\phi(T^*)||_{\ell^{2\sigma+2}}^{2\sigma+2}\nonumber\\
&\geq&\sum_{\kappa=1}^{N}||\nabla_k^+\phi(T^*)||^2_{\ell^2}-\lambda_*\sum_{\kappa=1}^{N}||\nabla_k^+\phi(T^*)||^2_{\ell^2}\\
&=&(1-\lambda_*)\sum_{\kappa=1}^{N}||\nabla_k^+\phi(T^*)||^2_{\ell^2}>0,
\end{eqnarray*}
under the assumption $\lambda_*<1$, justifying the first of the conditions on the initial data described in (\ref{glona}). The contradiction implies that $T^*=\infty$.
Now, since $\mathcal{K}(\phi(t)\geq 0$, for all $t\in [0,+\infty)$, the dissipation of energy identity (\ref{dissE}), implies that 
\begin{eqnarray*}
\frac{1}{2}||\dot{\phi}(t)||_{\ell^2}^2+\frac{1}{2}\sum_{\kappa=1}^{N}||\nabla_k^+\phi(t)||^2_{\ell^2}
&\leq& \mathcal{H}(0)+\frac{1}{2\sigma+2}||\phi(t)||^{2\sigma+2}_{\ell^{2\sigma+2}}\nonumber\\
&\leq& \mathcal{H}(0)+\frac{1}{2\sigma+2}\sum_{\kappa=1}^{N}||\nabla_k^+\phi(t)||^2_{\ell^2}.\nonumber
\end{eqnarray*}
Thus, we have that
\begin{eqnarray*}
\frac{1}{2}||\dot{\phi}(t)||_{\ell^2}^2+\frac{\sigma}{2\sigma+2}\sum_{\kappa=1}^{N}||\nabla_k^+\phi(t)||^2_{\ell^2}\leq\mathcal{H}(0),
\end{eqnarray*}
which implies the bound
\begin{eqnarray}
\label{es4}
\frac{1}{2}||\dot{\phi}(t)||_{\ell^2}^2+\frac{1}{2}\sum_{\kappa=1}^{N}||\nabla_k^+\phi(t)||^2_{\ell^2}\leq\frac{(\sigma+1)\mathcal{H}(0)}{\sigma}
,\;\;\mbox{for all}\;\;t\in [0,+\infty).
\end{eqnarray}
With the aim to estimate the quantity $\int_{0}^t K(\phi(s))ds$, we integrate once again (\ref{blo9}) with respect to time, to get the identity
\begin{eqnarray}
\label{es7}
(\dot{\phi}(t),\phi(t))_{\ell^2}&-&(\phi_0,\phi_1)_{\ell^2}-\int_{0}^{t}||\dot{\phi}(s)||_{\ell^2}^2ds
+\int_{0}^{t}\sum_{\kappa=1}^{N}||\nabla^+_{\kappa}\phi(s)||_{\ell_2}^2ds\nonumber\\
&+&\nu\int_{0}^{t}(\dot{\phi}(s),\phi(s))_{\ell^2}ds=\int_{0}^{t}||\phi(s)||^{2\sigma+2}_{\ell^{2\sigma+2}}.
\end{eqnarray}
Therefore, from (\ref{discmod}) and (\ref{es7}), we get
\begin{eqnarray}
\label{es8}
\;\;\;\;\;\int_{0}^{t}K(\phi(s))ds&\leq& \int_{0}^{t}||\dot{\phi}(s)||_{\ell^2}^2ds+|(\dot{\phi}(t),\phi(t))_{\ell^2}|
+ |(\phi_0,\phi_1)_{\ell^2}|\nonumber\\
&&+\nu\int_{0}^{t}|(\dot{\phi}(s),\phi(s))_{\ell^2}|ds.
\end{eqnarray}
By using (\ref{lp5}), and the bounds (\ref{blo93}) and (\ref{blo94}),  for the second and the last term of the rhs of (\ref{es8}), we obtain the estimates
\begin{eqnarray}
\label{es8a}
|(\dot{\phi}(t),\phi(t))_{\ell^2}|&\leq&\frac{1}{2}\left\{||\dot{\phi}(t)||^2_{\ell^2}+||{\phi}(t)||^2_{\ell^2}\right\}
\leq\frac{\mathcal{H}(0)+\rho_0}{2},\\
\label{es9}
\nu\int_{0}^{t}|(\dot{\phi}(s),\phi(s))_{\ell^2}|ds
&\leq&\nu\int_{0}^t||\dot{\phi}(s)||_{\ell^2}||\phi(s)||_{\ell^2}ds\nonumber\\
&\leq&\nu\left(\int_0^t||\phi(s)||^2_{\ell^2}ds\right)^{1/2}\left(\int_{0}^{t}||\dot{\phi}(s)||^2_{\ell^2}ds\right)^{1/2}\nonumber\\
&\leq&\nu\rho_0^{1/2}t^{1/2}
\left(\int_0^t||\dot{\phi}(s)||_{\ell^2}^2ds\right)^{1/2}
\nonumber\\
&\leq& (\nu\rho_0\mathcal{H}(0))^{1/2}t^{1/2}.
\end{eqnarray}
On the other hand, (\ref{dissE}) implies that
\begin{eqnarray}
\label{es10}
\frac{d}{dt}\left\{\mathcal{H}(\phi(t))(1+t)\right\}\leq \mathcal{H}(\phi(t)).
\end{eqnarray}
We shall integrate (\ref{es10}) with respect to time, and  we will insert the relation
\begin{eqnarray*}
-\mathcal{V}(\phi(t))=\frac{1}{2\sigma+2}K(\phi(t))+\frac{\sigma}{2\sigma+2}\sum_{\kappa=1}^{N}||\nabla^+_{\kappa}\phi(s)||_{\ell_2}^2,
\end{eqnarray*}
(obtained from (\ref{discmod}) and (\ref{Gal2})), to get the inequality
\begin{eqnarray}
\label{es11}
(1+t)\mathcal{H}(t)&\leq& \mathcal{H}(0)+\int_{0}^{t}\mathcal{H}(s)ds\nonumber\\
 &\leq&\mathcal{H}(0)+\frac{1}{2}\int_{0}^{t}||\dot{\phi}(s)||_{\ell^2}^2ds-\int_{0}^{t}\mathcal{V}(\phi(s))ds\nonumber\\
 &\leq&\mathcal{H}(0)+ \frac{1}{2}\int_{0}^{t}||\dot{\phi}(s)||_{\ell^2}^2ds+\frac{1}{2\sigma+2}\int_{0}^{t}K(\phi(s))ds
+\frac{\sigma}{2\sigma+2}\int_{0}^{t}\sum_{\kappa=1}^{N}||\nabla^+_{\kappa}\phi(s)||_{\ell_2}^2ds\nonumber\\
&\leq&\mathcal{H}(0)+\frac{1}{2}\int_{0}^{t}||\dot{\phi}(s)||_{\ell^2}^2ds+\frac{1}{2\sigma+2}\int_{0}^{t}K(\phi(s))ds
+\frac{\sigma}{\lambda^{*}(2\sigma+2)}\int_{0}^{t}K(\phi(s))ds.
\end{eqnarray}
where $\lambda^{*}=1-\lambda_*>0$. Finally, from (\ref{es11}), the estimates (\ref{blo93}) and  (\ref{es8})- (\ref{es8a})-(\ref{es9}), we obtain the energy decay estimate 
\begin{eqnarray}
\label{energydecay}
\mathcal{H}(t)&\leq& L_1\frac{1}{1+t}+L_2\frac{t^{\frac{1}{2}}}{1+t}.
\end{eqnarray}
The constants $L_1,L_2$ are found to be
\begin{eqnarray}
\label{1stcon}
L_1&=&\left(\frac{\lambda^{*}+\sigma}{2\lambda^*(\sigma+1)}+\frac{\nu+3}{\nu}\right)
\mathcal{H}(0)+\frac{\lambda^*+\sigma}{2\lambda^*(\sigma+1)}\left(\frac{\rho_0}{2}+|(\phi_0,\phi_1)_{\ell^2}|\right),\\
\label{2ndcon}
L_2&=&\frac{\lambda^{*}+\sigma}{2\lambda^*(\sigma+1)}(\nu\rho_0\mathcal{H}(0))^{1/2}.
\end{eqnarray}

Let us note e proof remains valid for the case $m>0$, under similar assumptions on the initial data, under slight modifications. \ \ $\diamond$
\begin{remark} 
\label{rem1}
{\em (Nonlinearity exponents). A review of the results in Section 3, implies that unique solutions exist locally (and globally under suitable conditions on the initial data), independently of  the growth of the nonlinearity and the dimension of the lattice. The same holds for the conditions on global nonexistence of solutions. This observation, although a simple consequence of (\ref{lp1}), the equivalence of norms of $\ell^2$ and of the ``auxiliary'' space of ``discrete derivatives'' $\ell^2_1$ (a discrete analogue of the Sobolev space $H^1(\mathbb{R}^N)$), looks interesting if one compares the corresponding results for the KG-partial differential equation (\ref{waveeq}) with $\mathrm{(P)}$.
We recall a representative result, stated in \cite[Section I, pg. 327]{Nakao93}, \cite{cazh}. It is well known that local existence depends on the growth of the nonlinearity and the dimension of the domain, since unique local solutions exist if $0<\sigma\leq 1/(N-2)$ and $0<\sigma<\infty$ only for the cases $N=1,2$,  a consequence of the Sobolev embedding theorems. Similar restrictions hold for the case of a bounded domain of $\mathbb{R}^N$.

However, in the global existence result of Section 4, for the linearly damped lattice, it appears a dependence not only on the size of the initial data but also on the nonlinearity exponent and the dimension of the lattice. This is not only in contrast with the DNLS lattice, but it also demonstrates an interesting difference with the continuum DKG partial differential equation: while in the continuum case the nonlinearity exponent is required to be smaller than a critical value in the case $N\geq 3$ ($\sigma\leq 1/(N-2)$), in the discrete case and for initial data satisfying (\ref{glona}), the nonlinearity exponent {\em has to be greater than  a critical value}-depending on the dimension of the lattice-$\sigma\geq 2/N$.
Thus, for global existence and energy decay of some initial data in the linearly damped DKG lattice, sufficient growth of the nonlinearity is  required. Considering for example the equilibrium solutions of the linearly damped DKG, and taking into account the results of \cite[Theorems 8 \& 9, pg. 1855]{PK} for the $1D$-lattice (see also \cite[Remark, pg. 1852]{PK}), the result of Theorem \ref{Nakono} seems to be also a verification of a Weinstein type result on higher dimensional lattices for the DKG, on the instability of nontrivial equilibrium solutions (and existence of excitation thresholds) for this type of solutions. 
\newline
{\em  (Generalizations)}. The arguments described in Sections 3 and 4, could be repeated for more general nonlinearities than the model case $\mathrm{(P)}$. We refer to those of 
\cite[Proposition 6.3.3, Proposition 6.4.1]{cazh} and those with the structure assumed in \cite{Pohoz,Pucci98}. For further differences with the continuous problem, we note as an example, that the proof of Theorem \ref{adaptcaz}, indicates that assumptions on the growth and the sign of the function $G(x)=\int_{0}^{x}F(s)ds$, as those of \cite[Proposition 6.3.3 \& Remark 6.3.4]{cazh} for $|x|$ small, do not appear in the discrete setting, for global existence of small solutions. 

The results hold also for DKG equations, assuming Dirichlet boundary conditions. However, the equivalence of norms in the finite dimensional spaces, allows for a transfer of the calculations of \cite{Ball1b}, to the finite dimensional  DKG equation, and to produce a finite time blow-up result (rather than global nonexistence of solutions). 

The assumptions on generalized discrete operators, which are not necessarily discretization of the Laplacian, described in \cite[pg. 347-348]{SZ2}, and which give rise to the their self-adjointness properties, and definition of appropriate equivalent norms \cite[pg. 349]{SZ2}, imply that the results of the previous sections, can be extended to DKG equations, involving such operators. 

Other, possibly interesting improvements, could consider  (a) the case of nonlinear damping. For this case, the references \cite{Geo94,Nakao93,Pucci98,Ono97}, should be valuable tools.  (b) Discretizations of reaction-diffusion equations \cite{Bates2, SZ2}, involving ``blow-up'' type nonlinearities.} 
\end{remark}  
\subsection{Forced solutions for parametrically damped and driven lattices and equilibria of the DKG equation.}
\label{rem2}
The short discussion is motivated by \cite[Section VI]{Yuri}, which considers the driven damped chain, of non.linear oscillators
\begin{eqnarray}
\label{Yuri1}
m\ddot{\phi}_n-\epsilon(\phi_{n-1}-2\phi_n+\phi_{n+1})+m\omega_0^2\sin \phi_n +\nu\dot{\phi}_n=f\cos(2\omega_e t)\sin\phi_n,\;\;n\in\mathbb{Z}.
\end{eqnarray}
Expanding  $\sin\phi_n$ in the Taylor series, one obtains a cubic DKG equation. The constant $\omega_0$, is the frequency of small amplitude vibrations in a well of a substrate potential, while $\omega_e$ denotes the frequency of the driving force $f$. When the force $f$ and the damping $\nu$ are small, one may try to look for {\em forced solutions} of the form
\begin{eqnarray}
\label{Yuri2}
\phi_n(t)=\psi_n\cos(\omega_e t+\Theta),\;\;n\in\mathbb{Z}.
\end{eqnarray}
Under the approximation described in \cite[pg. 3169]{Yuri}, the resulting equation for the real function $\psi_n$ describing the wave envelope, reads as
\begin{eqnarray}
\label{Yuri3}
-\epsilon(\psi_{n-1}-2\psi_n+\psi_{n+1})+\Omega\psi_n=\lambda\psi_n^3,\;\;n\in\mathbb{Z},
\end{eqnarray}
where the constants $\Omega,\lambda$, are given by
\begin{eqnarray*}
\Omega&=&\left(m(\omega_0^2-\omega_e^2)-2\epsilon-\frac{1}{2}f\cos(\Theta)\right),\\
\lambda&=&\frac{3}{4}\beta,\;\; \sin(2\Theta)=\frac{2\gamma\omega_e}{f},
\end{eqnarray*}
and $\Theta$ denotes the phase shift. 

The simple proofs of the results of \cite{AN, K1}, obviously can be applied for the rigorous justification of the existence and nonexistence of nontrivial forced solutions (equation (\ref{Yuri3}), is exactly the same with the one satisfied by the complex function $\psi_n$, in the ansatz for the breather solution $e^{i\Omega t}\psi_n$ of the DNLS equation). The same is valid, for the problem of nontrivial equilibria, of the DKG equation (\ref{lat1})-(\ref{lat2})-$\mathrm{(P})$, satisfying the equation
$$-(\Delta_d\phi)_n+m\phi_n=|\phi_n|^{2\sigma}\phi_n,\;\;n\in\mathbb{Z}^N.$$
\begin{theorem}
\label{Yuri4}
A.(Dirichlet Boundary Conditions-Finite lattice) Let $\epsilon>0$. For any $\Omega>0$ there exists a nontrivial solution of (\ref{Yuri1}).\newline
A1. Consider the DKG equation (\ref{lat1})-(\ref{lat2})-$\mathrm{(P})$. For any $m>0$, there exists a nontrivial equilibrium solution.\newline
B. (Dirichlet Boundary conditions or infinite lattice) Let $\epsilon>0$. There exist no nontrivial forced solution for (\ref{Yuri1}), of power less than 
\begin{eqnarray}
\label{disp2}
P_{\mathrm{min}}(\Omega,\lambda)=\frac{\Omega}{3\lambda}.
\end{eqnarray}
B.1 There exist no nontrivial equilibria  for (\ref{lat1})-(\ref{lat2})-$\mathrm{(P})$, of power less than
\begin{eqnarray}
\label{disp2a}
P_{\mathrm{min}}(m,\sigma)=\left(\frac{m}{2\sigma +1}\right)^{1/\sigma}.
\end{eqnarray}
\end{theorem}. 

The proof, is made by a discrete analogue of the Mountain Pass Theorem \cite{Amb}. It is valid in the case of a finite dimensional problem when $\lambda=\mathrm{const}$ \cite{AN}, and can be extended to infinite dimensional lattices when the anharmonic parameter is an element of an appropriate $\ell^p$-space \cite{K1}. The extension of the method in order to cover the case $\lambda=\mathrm{const}$ in infinite lattices possibly needs the concentration compactness arguments of \cite{Wein99}. 

On the other hand, the non-existence results-obtained by a fixed point argument- are valid for the infinite dimensional lattices, both for the case of site dependent and site independent anharmonic parameter. Moreover it provides a simple  proof on the existence of excitation thresholds both in the case of DNLS and DKG cases, \cite{Wein99, PK}. Alternatively to those proved in \cite{Wein99}-depending on the dimension of the lattice and the exponent of the nonlinearity)-the threshold (\ref{disp2}), depends on $\Omega$ or $m$ (or the frequency of the standing wave of DNLS equation) and the exponent of the nonlinearity, see \cite[Theorem 2.5]{K1}. 

Theorem \ref{Yuri4}B, is valid for any nonlinearity which defines a locally Lipschitz map on $\ell^2$, and similar thresholds can be obtained. Another interesting example is the so called saturable nonlinearity, $F(s)=\frac{s}{1+|s|^2}$. We refer to \cite{EilSatDNLS} for the saturable DNLS problem. 

In conclusion, the assumptions on the initial data needed for the results of Sections 3 \& 4, imply the instability of the non-trivial equilibria (or standing waves) for the DKG equation (\ref{lat1})-(\ref{lat2})-$\mathrm{(P})$, for such data.
\vspace{0.2cm}
\newline
{\bf Acknowledgments}. I would like to thank the referee for his valuable comments and suggestions (especially for the discussion on the linearly damped DKG), improving considerably the presentation of the manuscript. This work was partially supported by the  joint research project proposal ``Pythagoras I-Dynamics of Discrete and Continuous Systems and Applications''-NTUA and UOA.
\bibliographystyle{amsplain}

\end{document}